\documentclass{article}

\usepackage{PRIMEarxiv}
\usepackage[utf8]{inputenc}
\usepackage[T1]{fontenc}
\usepackage{hyperref}
\usepackage{url}
\usepackage{booktabs}
\usepackage{amsfonts}
\usepackage{nicefrac}
\usepackage{microtype}
\usepackage{fancyhdr}
\usepackage{fancyhdr}
\usepackage{graphicx}
\graphicspath{{media/}}

\title{AIoT-Based Smart Education System: A Dual-Layer Authentication and Context-Aware Tutoring Framework for Learning Environments.
\thanks{\textit{\underline{Citation}}: \textbf{Neelakantan, A., Satpute, P., Shinde, P., Devang, T. M. AIoT-Based Smart Education System: Enhancing Learning through Integrated Artificial Intelligence and Internet of Things.}}}

\author{
  Adithya Neelakantan\thanks{aneelaka@syr.edu; Equal contribution}, 
  Pratik Satpute\thanks{pvsatput@syr.edu; Equal contribution}, 
  Prerna Shinde\thanks{pgshinde@syr.edu; Equal contribution}, 
  Tejas Manjunatha Devang\thanks{tdevang@syr.edu; Equal contribution} \\
  College of Engineering and Computer Science, 
  Syracuse University \\
  Syracuse, NY
}

\begin{document}
\maketitle

\begin{abstract}
The AIoT-Based Smart Education System integrates Artificial Intelligence and IoT to address persistent challenges in contemporary classrooms – attendance fraud, lack of personalization, student disengagement, and inefficient resource use. The unified platform combines four core modules: (1) a dual-factor authentication system leveraging RFID-based ID scans and WiFi verification for secure, fraud-resistant attendance; (2) an AI-powered assistant that provides real-time, context-aware support and dynamic quiz generation based on instructor-supplied materials; (3) automated test generators to streamline adaptive assessment and reduce administrative overhead; and (4) the EcoSmart Campus module, which autonomously regulates classroom lighting, air quality, and temperature using IoT sensors and actuators. Simulated evaluations demonstrate the system’s effectiveness in delivering robust real-time monitoring, fostering inclusive engagement, preventing fraudulent practices, and supporting operational scalability. Collectively, the AIoT-Based Smart Education System offers a secure, adaptive, and efficient learning environment, providing a scalable blueprint for future educational innovation and improved student outcomes through the synergistic application of artificial intelligence and IoT technologies.
\end{abstract}

\keywords{AIoT \and Smart Classrooms \and RFID-based Attendance \and AI-Powered Chat bot \and EcoSmart Campus System \and Automated Quiz Generation \and WiFi Authentication}

\section{Introduction}
The rapid advancement of technology has ushered in a new space for educational innovation, driven largely by the convergence of Artificial Intelligence (AI) and the Internet of Things (IoT). There has been some work put into how to bring AI into objects around the house – from a HomePod that is an intelligent network hub that can manage different home functions like lights and thermostats, to wearable devices that are powered by AI, like watches that observe health patterns and help people on the go. Bringing this into the education sector, and finding the right gaps to bridge within the traditional classroom setting has proven to be largely relevant. This amalgamation, commonly referred to as the Artificial Intelligence of Things (AIoT), has enabled the creation of intelligent environments where devices are not only interconnected but also capable of learning from and autonomously responding to real-time data streams. In the educational domain, the adoption of AIoT has promised to dramatically enhance the learning experience by enabling smart classrooms – spaces that adapt dynamically to students’ needs through personalized content delivery, advanced analytics on engagement, secure automated attendance, and intelligent environmental controls.

While both AI and IoT technologies have been individually applied in educational settings, such as AI-powered learning platforms and IoT-enabled environmental monitoring, fully integrated AIoT systems remain uncommon. The primary challenge of this prototype lies not only in developing sophisticated AI algorithms or deploying diverse IoT sensor networks, but in the seamless orchestration of these components. Achieving reliable interoperability, real-time data sharing, security, and actionable analytics in a unified ecosystem is a major technical hurdle. Issues like the lack of standard protocols, difficulties in data management and synchronization, and concerns over privacy and scalability have all contributed to the relative rarity of holistic AIoT deployments in today's schools and universities.

As a result, most educational solutions leverage these technologies in isolation, offering either smart attendance, digital assessment platforms, or automated classroom management, but rarely an integrated, context-aware, and adaptive environment. Bridging this gap is vital for realizing the full educational potential of digital transformation. The presented work aims to address this communication challenge between components by proposing and prototyping an AIoT-based platform that not only brings together advanced AI and IoT subsystems, but does so in a way designed for coherence, extensibility, and practical deployment in real-world learning spaces.

\section{Significance}
This research contributes to the ongoing evolution of educational technology by demonstrating the practical integration of AI and IoT in classroom environments. The proposed AIoT-based Smart Education System makes use of advanced components – RFID authentication, WiFi validation, real-time sensor-actuator networks to monitor environment optimization, AI-driven chat bots, and automated quiz generation – all to establish a cohesive, adaptive, and scalable architecture where the different subparts are communicating with each other. By tackling interoperability and modularity, the system is able to advance beyond individual AI or IoT solutions, offering an extensible platform that can accommodate future innovations and deployments by amalgamating both technologies.

Importantly, the system aims to solve the challenges that have hindered educational efficiency and equity. Truancy, student disengagement, lack of personalized learning support, and inefficient use of classroom resources are well-documented obstacles in both early and higher education settings. The features provided within this research prototype foster inclusivity and real-time response, leading to improved student participation and personalized academic support. Automated classroom management not only enhances operational efficiency, reduces the administrative burden for educators, but also promotes data-driven insights into both individual and group progress.

By aligning education with the digital trend cycles and workforce demands of the larger technological scene, this work helps prepare students with in-demand academic knowledge. The adaptability of the system to diverse learning needs and its potential for remote or hybrid implementations render it particularly relevant in a post-pandemic world, where flexible and tailored digital learning is paramount.

The AIoT Smart Education System presented within this work creates an engaging and secure framework for next-generation classrooms and campuses. Unlike prior AIoT systems, this work introduces a fully unified, co-existing platform that tightly integrates dual-layer RFID+WiFi authentication, real-time AI-powered tutoring trained on classroom-specific materials, adaptive quiz generation, and IoT-driven environmental optimization—all within a single, interoperable framework. Whereas most existing approaches address these dimensions in isolation, our system demonstrates robust orchestration and secure access control between all modules, allowing for personalized support and management at both the individual and classroom scale. This holistic architecture establishes a newer benchmark for modularity and practical deployment in a real-world educational environment.

\section{Problem Analysis}
Traditional classroom environments continue to face constant challenges that hinder both teaching effectiveness and student engagement. One of the primary issues is unreliable and easily manipulated attendance systems. Manual roll calls or single-layer authentication methods are vulnerable to fraud, and students often mark attendance on behalf of their peers. This is not only unfair to those who diligently attend classes, but it also diminishes the value of physical attendance. Many students, especially those who are introverted or lack confidence, may hesitate to ask questions during lectures. This could lead to students falling behind on several key concepts.

Also, the lack of personalized educational support means that students receive generalized instruction, failing to deliver to individual learning speeds and knowledge gaps. This inability to offer real-time concept clarification or additional explanations during class often leads to missed learning opportunities.

Together, these issues call for an intelligent and integrated system that can ensure attendance integrity, reduce the administrative burden on teachers, support student learning in real time, and create an optimal learning environment. This prototype system addresses these concerns through a multifaceted approach involving RFID and WiFi-based attendance validation, AI-powered learning assistants, automated quiz generation, and IoT-driven environmental monitoring. 

\section{Methodology}
The project integrates four features — two IoT-driven modules implemented via Wokwi simulation (RFID-based authentication and the EcoSmart School system), and two AI-based modules (an automated Test/Quiz Generator and a context-aware classroom assistant chat bot). The main objective is to systematically handle widespread challenges encountered by both students and educators in modern learning environments and to empirically evaluate improvements in engagement, personalization, and administrative efficiency.

For example, introverted students can clarify doubts comfortably using the chat bot, which has been trained not only on internet data but also on educators' slide decks and uploaded learning materials. Attendance can be authenticated using a two step verification system that combines RFID scanning with WiFi-based validation through the school network. Students can test their knowledge and fill learning gaps with AI-generated quizzes, while teachers can get mixed bags of questions using the Test Generator to create balanced assessments from class materials. Also, classroom conditions can be optimized using sensor-actuator systems that monitor and adjust the environment through the EcoSmart infrastructure. 

The methodology uses modular development, each subsystem is developed and functionally validated individually before integration into the unified framework. This helped ensure scalability and to evaluate each subsystem’s function separately and as a unit.

The project consists of four major modules:
\begin{itemize}
    \item \textbf{RFID+WiFi Attendance}: Dual-factor student authentication utilizes RFID scans in conjunction with device-based WiFi verification. The combined approach ensures high integrity in attendance reporting and minimizes the potential for fraudulent check-ins.
    
    \item \textbf{AI-Powered Classroom Assistant}: Leveraging advances in AI, this subsystem provides personalized, lecture-specific query resolution. Professors upload instructional content, which is used to contextualize responses from the chat bot, supporting self-paced and inclusive learning—particularly beneficial for students reluctant to participate in group settings.
    
    \item \textbf{IoT EcoSmart System}: Deployment of ESP32 controllers and environmental sensors (including temperature, humidity, light, and air quality), this module enables continuous monitoring and active adjustment of classroom conditions. Real-time sensor data processed to maintain optimized learning environments using actuator controls.
    
    \item \textbf{AI Test Generator}: With Google’s Gemini API and retrieval-augmented generation models, this command-line tool automates the creation of multiple-choice assessments aligned with course topics. It allows for objective formative evaluation while supporting personalized remediation.
    
\end{itemize}

Each module is developed and validated separately and then integrated for full-system simulation. This research has helped us create a simulation for developing, integrating, and testing AIoT-enabled educational solutions in school settings.

\section{Implementation}
\subsection{System Architecture and Integration}
The AIoT Education System is structured as a dynamic and interoperable platform, bringing together edge IoT devices, machine learning models, and interactive interfaces. The core subsystems were developed as independent modules, with clearly defined data flow protocols, and were subsequently integrated to ensure robust end-to-end operation in both simulation and prototype environments.

\begin{figure}
    \centering
    \includegraphics[width=0.75\linewidth]{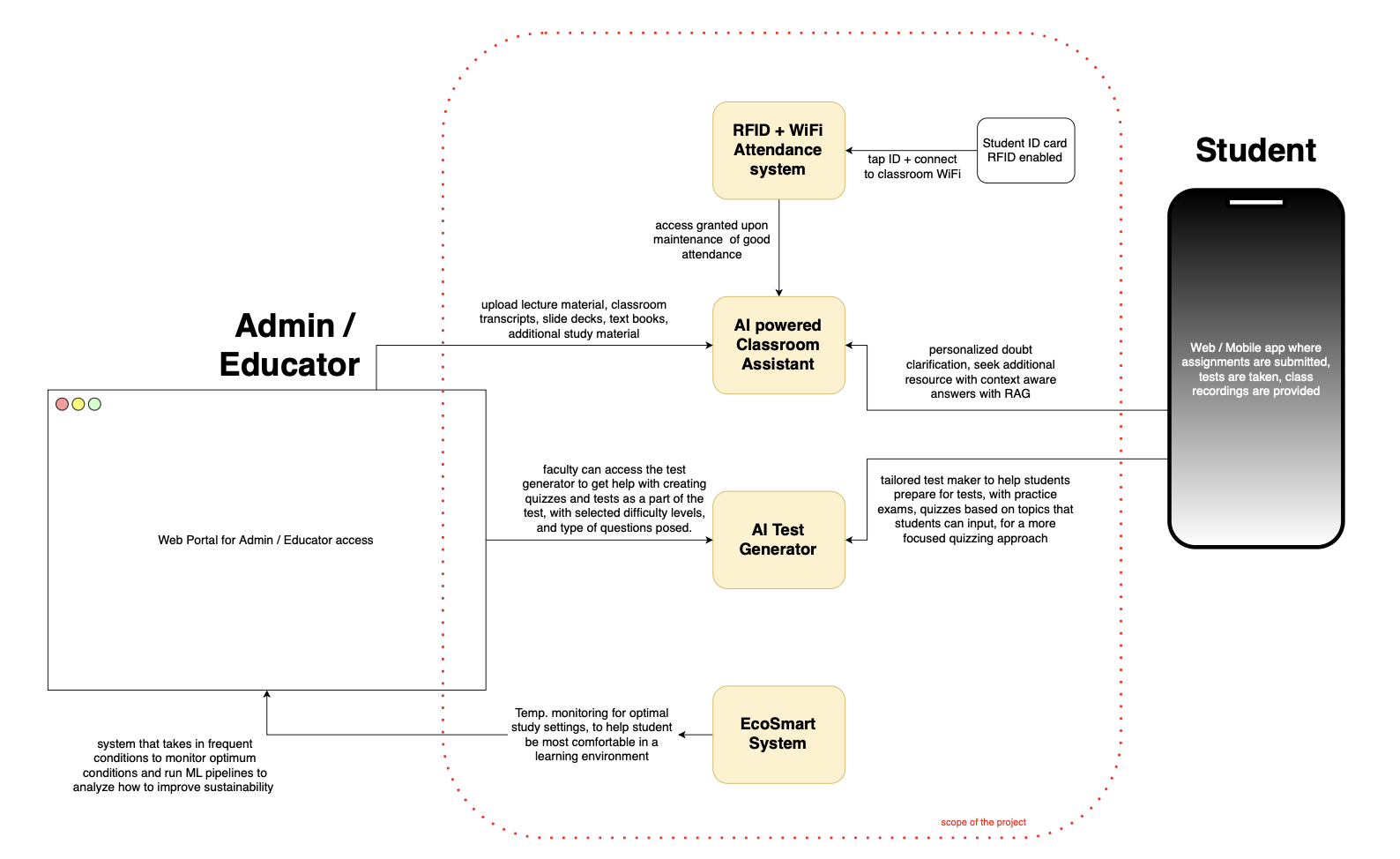}
    \caption{System architecture of the AIoT-Based Smart Education System. Dotted lines indicate planned future capability not yet implemented in current prototype.}
    \label{fig:system architecture}
\end{figure}

\subsection{Dual-Layer Attendance Verification}
The attendance module uses RFID and WiFi-based verification to secure classroom entry and prevent attendance fraud. On arrival, each student presents an RFID-enabled ID card at the entry sensor node. The ESP32 microcontroller reads the RFID tag and simultaneously validates the student’s device MAC address or login status on the local WiFi network. Both authentications are required within a session time window for positive attendance marking. Real-time responses are fed back to both the student (OLED display) and the instructor dashboard. The backend supports asynchronous error logging and retry mechanisms in cases of failed authentications for reliability. In the simulation phase, RFID functionality was successfully substituted with edge-triggered pushbutton logic to prototype the credentialing workflow, as supported RFID simulation is not supported on Wokwi.

\subsection{AI-Powered Classroom Assistant}
Professors upload instructional content (PDF slides, notes, transcripts) ahead of class using a web interface built with Streamlit. The ingestion pipeline parses these materials with pypdf, segments them using LangChain’s recursive character splitter, and generates vector embeddings via OpenAI models. All document vectors are indexed locally using FAISS, enabling semantic retrieval based on student queries. When a student submits a natural language question, the assistant performs a Similarity Search, retrieving the top 'k' relevant passages. It uses OpenAI’s generative API to format pedagogically newer and context-informed responses. Session-level caching ensures rapid repeat access and prevents unnecessary reprocessing. Access is gated to students verified by the dual-authentication mechanism, ensuring confidentiality and engagement only from physically present participants.

\subsection{AI Test and Quiz Generator} 
The assessment module allows instructors to generate adaptive, course-aligned quizzes using a pipeline built atop Google Gemini’s LLM and the Embedchain framework. The system is primarily composed of two Python files: ‘app.py’ (the main entry point) and ‘rag.py’ (main logic using Retrieval Augmented Generation, RAG). The user runs the program via the command line, providing a file path and optionally a topic and number of questions to generate.

The 'rag.py' module is used for document ingestion and interaction with the LLM. The main function ‘generateQuestions(numQuestions, topic)’ forms a natural language prompt and then queries the LLM. It makes sure that each question has a single correct answer and avoids markup formatting. The ‘app.py’ file handles command-line arguments and controls the interaction loop. It uses ‘argparse’ to accept the document path, subject/topic area to generate questions on and the number of questions. Once embedded, app.py enters a loop where the user types a topic (e.g., ‘MQTT’) and gets questions generated by the pre-trained model. The system uses Google’s Gemini Pro LLM and embedding APIs from the ‘google.generativeai’ library.

\subsection{IoT-Driven Environmental Optimization (EcoSmart Campus System)}
Environmental monitoring is developed in simulation using the Wokwi platform, with an ESP32 microcontroller orchestrating sensor readings (DHT22 for temperature/humidity, LDR for luminosity, MQ2 for air quality). The control logic is partitioned into acquisition (periodic polling), processing (threshold and hysteresis filtering), and actuation (activation of HVAC, lighting, and ventilation via relays/LEDs). Sensor calibration functions translate raw analog and digital values into human-interpretable metrics, supporting on-the-fly rescaling as classroom needs evolve. The OLED interface provides situational awareness for in-room users, while test routines iterate environmental parameters through all plausible operational ranges to validate system responses, logic stability, and error handling.

\section{Results and Discussion}

\subsection{EcoSmart Campus System}
\begin{figure}[h]
    \centering
    \includegraphics[width=0.6\linewidth]{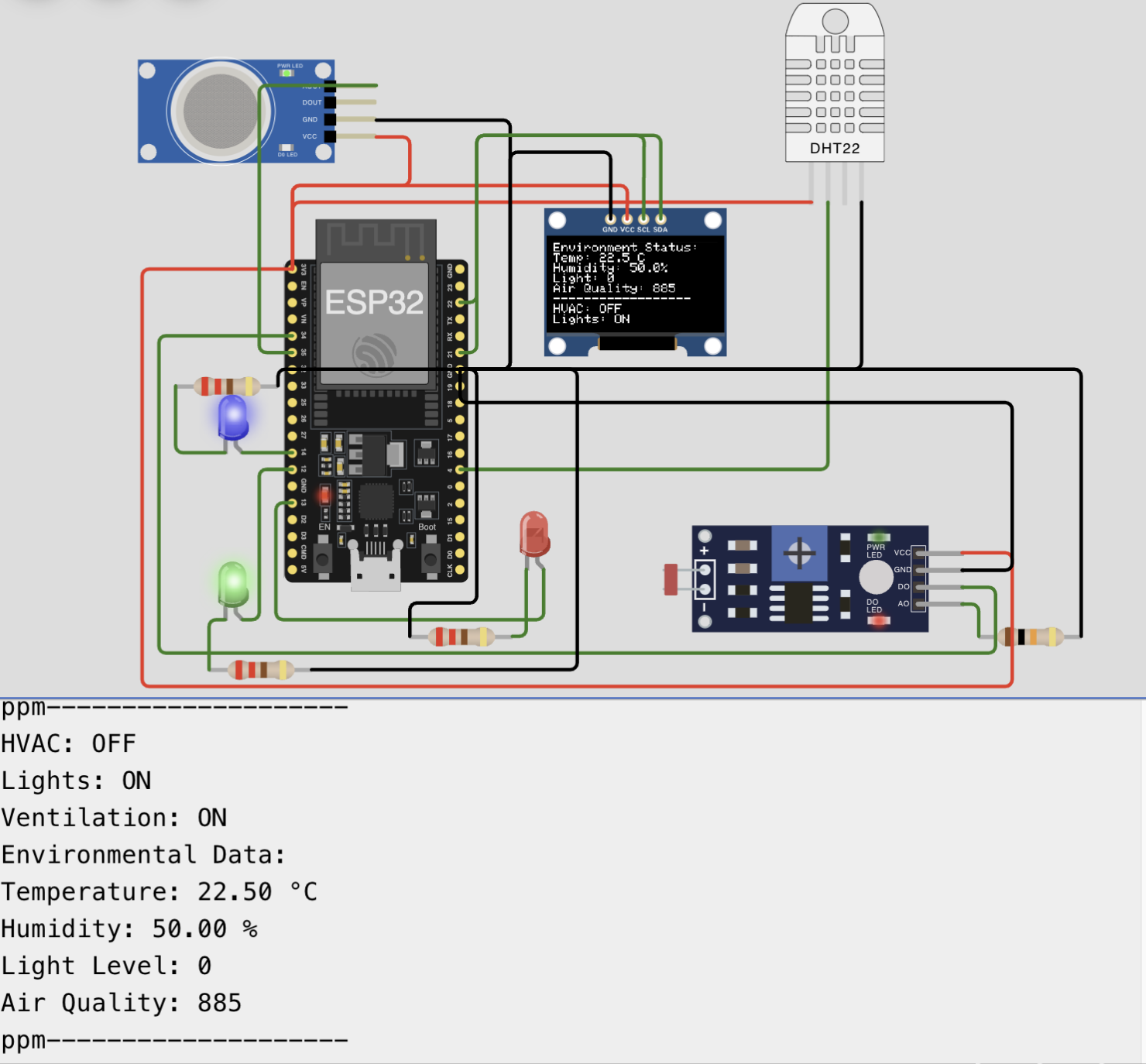}
    \caption{Simulation environment: IoT-based EcoSmart Campus System. Real-time sensor readings (temperature, humidity, light, air quality), actuator status (HVAC, lighting, ventilation), and live environmental data processed by the ESP32 controller.}
    \label{fig:ecosmart}
\end{figure}

The EcoSmart Campus System successfully validated the feasibility of applying IoT-driven environmental monitoring and adaptive control within instructional spaces. Simulation trials demonstrated consistent detection and quantification of temperature, humidity, ambient light, and air quality perturbations using appropriately calibrated sensors (DHT22, LDR, MQ2). Actuation logic (e.g., HVAC, lighting, ventilation) was consistently triggered based on predefined environmental thresholds, and all status changes were observable via OLED and LED indicators.

Iterative testing on the Wokwi platform minimized design-cycle overhead and enabled prompt prototyping, debugging, and software refinement. The control system was able to reliably prevent excessive on/off cycling (via hysteresis), and conversion functions accurately mapped sensor values to human-readable metrics. Despite simulation constraints, the architecture facilitated robust decision-making and paved the way for straightforward hardware deployment. Known limitations include variability in sensor response times, inability to simulate long-term material degradation, and limits in real-world spatial distribution. Compensatory design features included adaptive threshold logic and robust error handling to provide future extensibility and operational resilience in physical classroom environments.

\subsection{AI-Powered Classroom Assistant}
The classroom assistant module was deployed as a Streamlit-based web application employing a hybrid of document parsing (pypdf), AI embedding (OpenAI) and semantic search (FAISS) for real-time, context-aware tutoring aid. Instructor uploaded materials were semantically indexed before each session, and students were able to pose a broad array of questions, ranging from basic conceptual clarifications to complex contextual inquiries and questions regarding specific things said in class with the added context cues from transcripts being uploaded. 

Testing confirmed that the system delivered timely, relevant, and pedagogically aligned responses, with embedding reuse ensuring really low latency and high availability across several repeated queries. Usability feedback also informed on ease of interaction and accessibility, particularly for those students hesitant to participate in classroom settings. The assistant effectively supported self-paced and differentiated learning, while also reducing the instructor's burden through automated query resolution. 

\begin{figure}[h]
    \centering
    \includegraphics[width=0.5\linewidth]{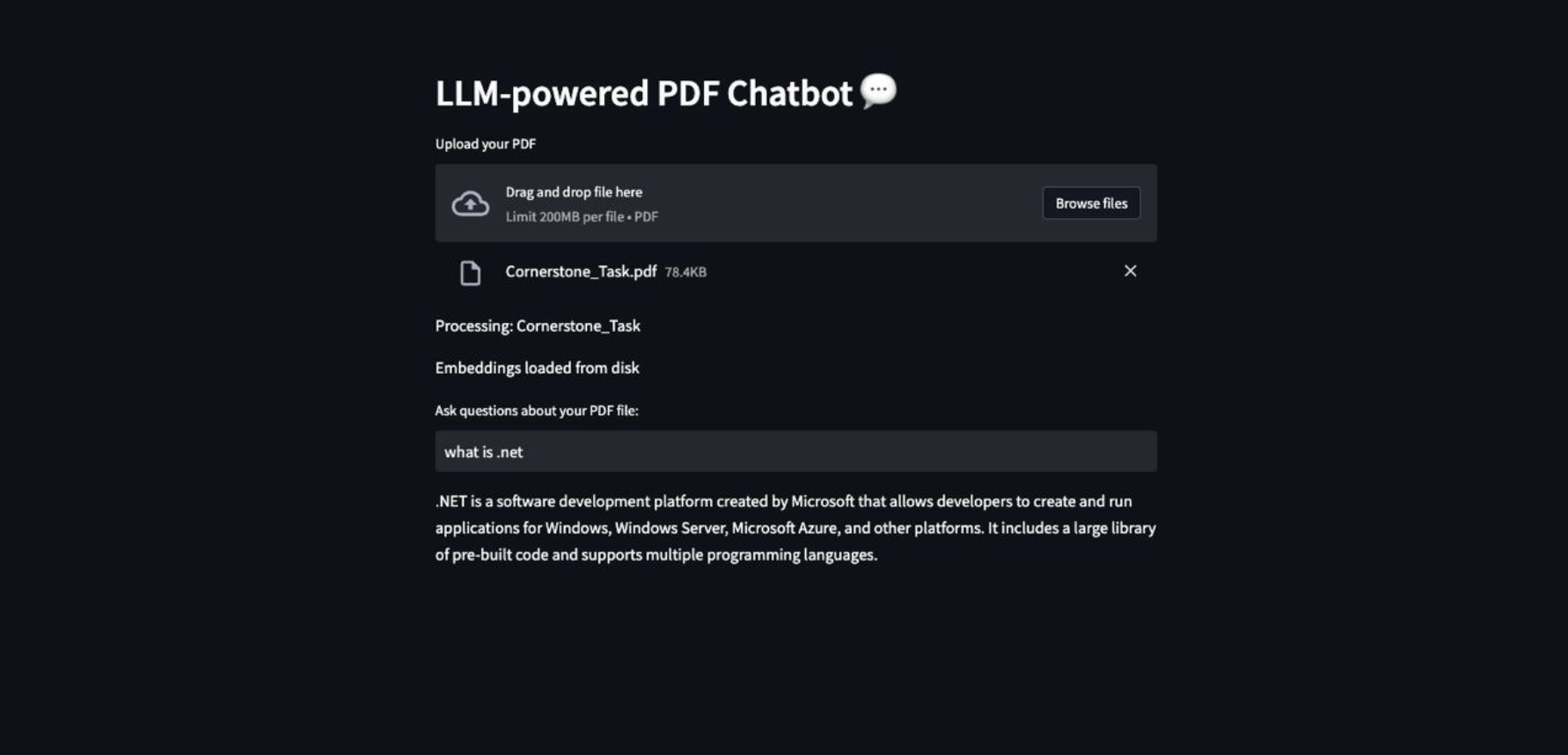}
    \caption{Screenshot of the LLM-powered PDF Chatbot interface, demonstrating real-time question answering on uploaded PDF documents.}
    \label{fig:placeholder}
\end{figure}

\subsection{Attendance Authentication System}
Simulation confirmed the reliability and usability of the attendance authentication system. During on-device testing within Wokwi, the ESP32 controller consistently connected to the correct WiFi network, and interactive prototypes displayed session readiness and state changes via OLED and serial logs. Simulated RFID events (with a push button interface) were detected accurately, with real-time feedback confirming attendance status and system readiness for subsequent inputs.

Multiple trial runs documented device coordination, accurate edge and debounce detection logic, and session logging. The modular software architecture accommodates expansion to fully integrated RFID hardware and cloud-based attendance record management. System validation had demonstrated strong potential for secure and fraud-resistant deployment in real educational environments, provided that appropriate hardware support and institutional adoption are in place.

\begin{figure}[h]
    \centering
    \includegraphics[width=0.5\linewidth]{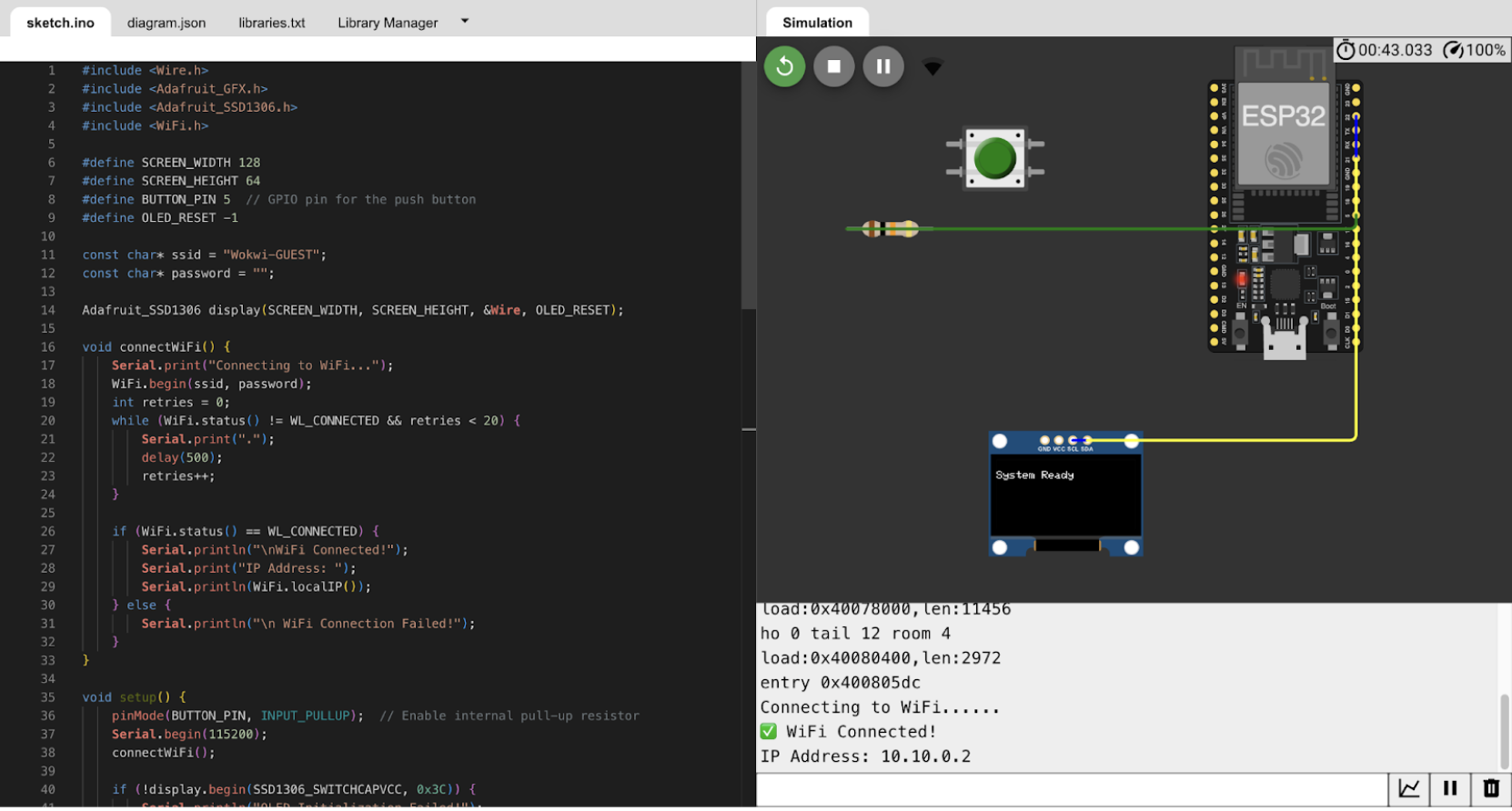}
    \caption{Wokwi platform simulating two level attendance authentication system - The ESP32 microcontroller, OLED status display, and pushbutton emulate RFID and WiFi-based check-in.}
    \label{fig:wokwi1}
\end{figure}

\begin{figure}[h]
    \centering
    \includegraphics[width=0.5\linewidth]{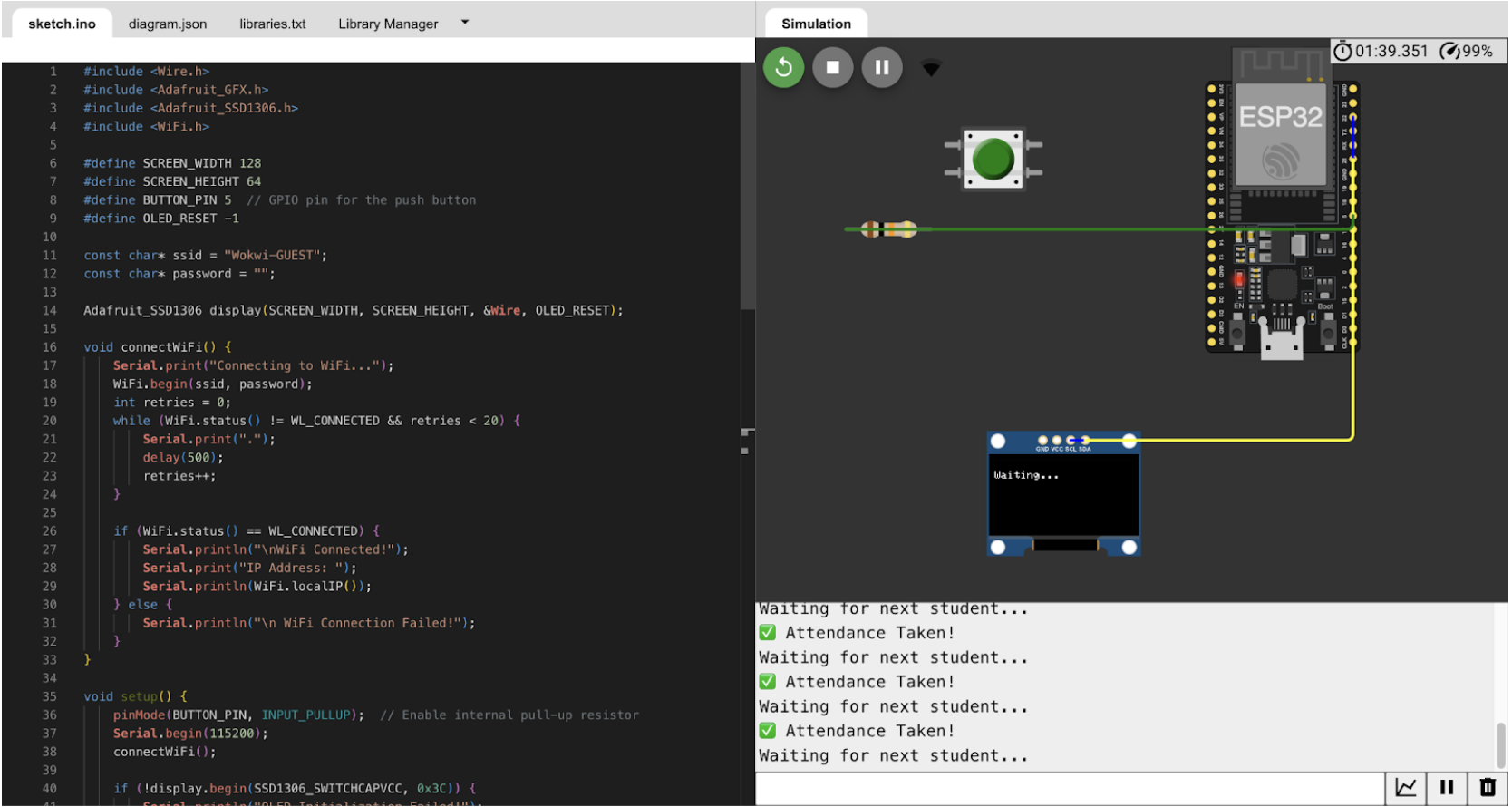}
    \caption{Simulated attendance workflow: Device status messages ('System Ready', 'Attendance Taken!'), WiFi connection feedback, and state transitions upon student interaction.}
    \label{fig:wokwi2}
\end{figure}

\subsection{AI Test Generator}
The quiz generator module, built with Google Gemini and Embedchain, was developed on controlled command-line interfaces. Input documents were parsed, queried for targeted topics (like Edge Node Functionality), and an adaptive set of five multiple-choice questions (default set to five for the scope of testing) was automatically generated per user session based on topics students wrote in. The model successfully applied retrieval-augmented generation, delivering varied and curriculum-aligned assessments.

These results were evaluated for accuracy and difficulty stratification, with configurable parameters for instructor control. The pipeline enabled iterative quiz revision during session runtime, ideally for student's use for practice sessions, and for faculty's use as an additional source to create tests with. The simulation confirmed a consistent parsing of PDF source materials and API-driven prompt composition. The image below shows an example with ‘Edge Node Functionality’ as a prompt from the related pdf to generate questions. It was able to parse the uploaded document and generate questions with different complexities.

\begin{figure}[h]
    \centering
    \includegraphics[width=0.5\linewidth]{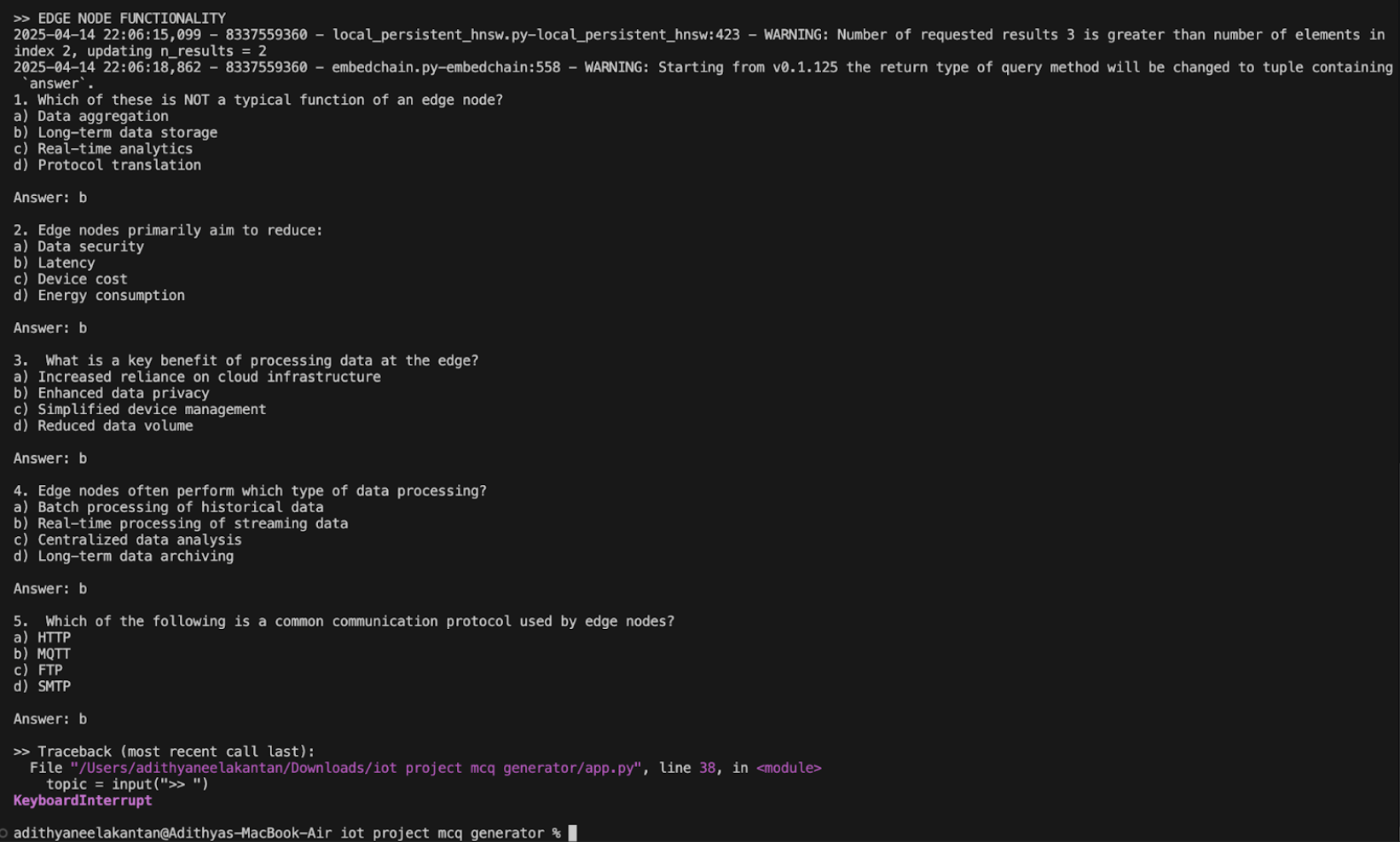}
    \caption{Command-line AI Test Generator module - Shown: dynamically-produced questions on ‘Edge Node Functionality’.}
    \label{fig:aitestgenerator}
\end{figure}

\section{Conclusion}
The AIoT Smart Education System addresses longstanding inefficiencies in traditional classrooms through the seamless integration of artificial intelligence and IOT technologies. By implementing dual-layer attendance authentication, automated adaptive assessment, real-time personalized AI tutoring, and IoT-based environmental controls, the platform empowers teachers, supports diverse students, and streamlines classroom management. Each module — attendance, tutoring, test generation, and environmental optimization — demonstrated its reliability and responsiveness in simulation, providing robust solutions to well-documented challenges such as participation, engagement, equity, and administrative overhead. The unified ecosystem fosters inclusive, data-driven, and classroom driven learning environments, enabling students to benefit from personalized support while encouraging active participation. Teachers would ideally be able to focus on instruction with reduced manual dependencies and students would gain access to actionable analytics that can further individualize the learning journey.

Simulation results have indicated strong technical feasibility and promise, but for real-world deployment, it would require further validation of scalability, security, privacy, and long-term reliability. Future research will focus on large-scale institutional pilots, integration with cloud learning platforms, privacy-preserving AI models, and expanded multi-modal sensor arrays. Ongoing evaluation will also explore the broader impacts on accessibility, equity, and motivation—critical issues for modern education systems.

Furthermore, as shown in the system architecture diagram, the project is limited in scope, studying how an integrated system could be created to combine the benefits of AI and IOT within an ecosystem. For it to be used at full performance capacity, it would require a robust web / application interface that is easily operable by both students and educators. It could be expanded into a larger system, one where several key players (like an ML pipeline to aim toward sustainability, a one-stop portal for students and teachers to access for additional resources, assignments, learning aids) are contributing toward a singular focus - providing the best resources for anyone willing to learn. 

Ultimately, this research lays the groundwork for next-generation smart campuses and adaptive learning environments, demonstrating that AIoT integration can deliver transformative improvements for both teaching and learning. By bridging digital and physical context, and prioritizing inclusivity alongside technical innovation, the proposed ecosystem sets a precedent for future studies in educational technology and learning sciences.

\end{document}